\documentstyle[12pt,aaspp4,flushrt]{article}

\def\eq#1{equation~(\ref{eq:#1})}
\def\eqs#1{equations~(\ref{eq:#1})}
\def\eqss#1{(\ref{eq:#1})}

\def\be{\begin{equation}}
\def\ee{\end{equation}}
\def\capt{\small \baselineskip 12pt }

\def\today{\ifcase\month\or January\or February\or March\or April\or May\or 
June\or July\or August\or September\or October\or November\or December\fi
  \space\number\day, \number\year}

\def\etal{{\it et al.\ }}

\def\eg{{e.g.}}

\def\ltsima{$\; \buildrel < \over \sim \;$}
\def\lsim{\lower.5ex\hbox{\ltsima}}
\def\gtsima{$\; \buildrel > \over \sim \;$}
\def\gsim{\lower.5ex\hbox{\gtsima}}
\def\ga{\mathrel{\hbox{\rlap{\hbox{\lower4pt\hbox{$\sim$}}}\hbox{$>$}}}}
\def\la{\mathrel{\hbox{\rlap{\hbox{\lower4pt\hbox{$\sim$}}}\hbox{$<$}}}}

\def\ifm#1{\relax\ifmmode#1\else$\mathsurround=0pt #1$\fi}

\def\hmpc{\,h\ifm{^{-1}}{\rm Mpc}}

\def\ln{{\rm ln}}

\def\sin{{\rm sin}}

\def\pa{\partial}
 
\def\la{\langle} \def\ra{\rangle}
\def\prop{\propto}

\def\solar{\ifmmode_{\mathord\odot}\;\else$_{\mathord\odot}\;$\fi}
\def\msun{{\rm M}_{\solar}}

\def\pmb#1{\setbox0=\hbox{#1}%
 \kern-.025em\copy0\kern-\wd0
 \kern.05em\copy0\kern-\wd0
 \kern-.025em\raise.0433em\box0}

\def\vv{\pmb{$v$}}
\def\vx{\pmb{$x$}}

\def\vnabla{\pmb{$\nabla$}}
\def\div{\vnabla\!\cdot\!}

\def\divv{\div\vv}

\def\del{\delta}
\def\delg{g}
\def\delgs{g_{\rm s}}
\def\sig{\sigma}
\def\sb{\sigma_{\rm b}}
\def\sg{\sigma_{\rm g}}

\def\sgs{\sigma_{\rm g,s}}
\def\eps{\epsilon}

\def\cor{{\tt r}}
\def\bh{\hat b}
\def\bt{\tilde b}
\def\bv{b_{\rm var}}
\def\bn{b_{\rm neg}}
\def\bp{b_{\rm pos}}
\def\av#1{\la #1 \ra}
\def\tg{\tilde g}
\def\td{\tilde\delta}
\def\Sb{S_{\rm b}}

\def\binv{b_{\rm inv}}
\def\betav{\beta_{\rm var}}
\def\betainv{\beta_{\rm inv}}
\def\betah{\hat\beta}

\def\done{\delta_1}
\def\dtwo{\delta_2}

\def\xigg{\xi_{\rm gg}}
\def\xigm{\xi_{\rm gm}}
\def\ximm{\xi_{\rm mm}}
\def\xibb{\xi_{\rm \eps \eps}}
\def\xibm{\xi_{\rm \eps m}}
\def\xb{r_{\rm b}}
\def\Pgg{P_{\rm gg}}
\def\Pgm{P_{\rm gm}}
\def\Pmm{P_{\rm mm}}

\begin{document}
\hfill\today
\medskip

\title{STOCHASTIC NONLINEAR GALAXY BIASING}

\author{Avishai Dekel \altaffilmark{1} and Ofer Lahav \altaffilmark{1,2}}
\authoremail{dekel@astro.huji.ac.il, lahav@ast.cam.ac.uk}

\altaffiltext{1}{Racah Institute of Physics, The  Hebrew University,
Jerusalem 91904, Israel}

\altaffiltext{2}{Institute of Astronomy, Madingley Road, Cambridge CB3 0HA, UK}


\begin{abstract}

We propose a general formalism for galaxy biasing and apply it to methods 
for measuring cosmological parameters, such as regression of light versus 
mass, the analysis of redshift distortions, measures involving skewness 
and the cosmic virial theorem.  The common linear and deterministic 
relation $g\!=\!b \del$ between the density fluctuation fields of 
galaxies $g$ and mass $\del$ is replaced by the conditional distribution 
$P(\delg\vert\del)$ of these random fields, smoothed at a given scale
at a given time.
The nonlinearity is characterized by the conditional mean 
$\av{\delg\vert\del}\!\equiv\!b(\del)\,\del$, while the local scatter 
is represented by the conditional variance $\sb^2(\del)$ and higher moments.  
The scatter arises from hidden factors affecting galaxy formation
and from shot noise unless it has been properly removed.
 
For applications involving second-order local moments, the biasing is 
defined by three natural parameters: the slope $\bh$ of the regression 
of $g$ on $\del$, a nonlinearity $\bt$, and a scatter $\sb$. The ratio 
of variances $\bv^2$ and the correlation coefficient $\cor$ 
mix these parameters.
The nonlinearity and the scatter lead to underestimates of order 
$\bt^2/\bh^2$ and $\sb^2/\bh^2$ in the different estimators of $\beta$ 
($\sim \Omega^{0.6}/\bh$).  The nonlinear effects are typically smaller.

Local stochasticity affects the redshift-distortion analysis only by
limiting the useful range of scales, especially for power spectra. 
In this range, for linear stochastic biasing, the analysis reduces to Kaiser's 
formula for $\bh$ (not $\bv$), independent of the scatter.  The distortion 
analysis is affected by nonlinear properties of biasing but in a weak way. 

Estimates of the nontrivial features of the biasing scheme are made based 
on simulations and toy models, and strategies for measuring them are 
discussed.  They may partly explain the range of estimates for $\beta$.

\end{abstract}

\subjectheadings{
cosmology: theory --- 
cosmology: observation ---
dark matter --- 
galaxies: distances and redshifts ---
galaxies: formation --- 
galaxies: clustering ---
large-scale structure of universe}


\section{INTRODUCTION}
\label{sec:intro}

Galaxy ``biasing" clearly exists.    
The fact that galaxies of different types cluster differently 
(\eg, Dressler 1980; Lahav, Nemiroff \& Piran 1990, 
Santiago \& Strauss 1992; Loveday \etal 1995;
Hermit \etal 1996; Guzzo \etal 1997) 
implies that not all of them are exact tracers of the underlying 
mass distribution.
It is obvious from the emptiness of large voids (\eg, Kirshner \etal 1987)
and the spikiness of the galaxy distribution with $\sim 100\hmpc$
spacing (\eg, Broadhurst \etal 1992), especially at high redshifts 
(Steidel \etal 1996; 1998), that if the structure has evolved by standard 
gravitational instability theory (GI) then the galaxy distribution 
must be biased.

Arguments for different kinds of biasing schemes have been put forward
and physical mechanisms for biasing have been proposed
(\eg, Kaiser 1984; Davis \etal 1985; Bardeen \etal 1996; 
Dekel \& Silk 1986; Dekel \& Rees 1987; 
Braun, Dekel \& Shapiro 1988;
Babul \& White 1991; Lahav \& Saslaw 1992).
Cosmological simulations of galaxy formation clearly indicate galaxy biasing,
even at the level of galactic halos
(\eg, Cen \& Ostriker 1992; Kauffmann, Nusser \& Steinmetz 1997;
Blanton \etal 1998; Somerville \etal 1998). 
The biasing becomes stronger at higher redshifts
(\eg, Bagla 1998a, 1998b; Jing \& Suto 1998; Wechsler \etal 1998).

The biasing scheme is interesting by itself as a constraint
on the process of galaxy formation,
but it is of even greater importance in many attempts to estimate
the cosmological density parameter $\Omega$.
If one assumes a linear and deterministic biasing relation of the sort
$\delg=b\del$ between the density fluctuations of galaxies and mass, 
and applies the linear approximation for gravitational instability,
$\divv = -f(\Omega) \del$ with $f(\Omega)\approx\Omega^{0.6}$ 
(\eg Peebles 1980),
then the observables $\delg$ and $\divv$ are related via the
degenerate combination $\beta\equiv f(\Omega)/b$.
Thus, one cannot pretend to have determined $\Omega$ by measuring $\beta$
without a detailed knowledge of the relevant biasing scheme.

It turns out that different methods lead to different estimates of $\beta$,
sometimes from the same data themselves 
(for reviews see Dekel 1994, Table 1; Strauss \& Willick 1995, Table 3;
Dekel, Burstein \& White 1997; Dekel 1998a).
Most recent estimates for optical and IRAS galaxies
lie in the range $0.4 \leq \beta \leq 1$. 
The methods include, for example:
(a) comparisons of local moments of $g$ (from redshift surveys) and $\del$
(from peculiar velocities) or the corresponding power spectra or 
correlation functions; 
(b) linear regressions of the fields 
$g$ and $\del$ or the corresponding velocity fields;
(c) analyses of redshift distortions in redshift surveys;
and
(d) comparisons of the Cosmic Microwave Background (CMB) dipole with 
the Local-Group velocity as predicted from the galaxy distribution.

In order to sharpen our determination of $\Omega$ it is important
that we understand the sources for this scatter in the estimates of 
$\beta$.  Some of this scatter is due to the different
types of galaxies involved, and some may be due to unaccounted-for effects of 
nonlinear gravity and perhaps other sources of systematic errors in the 
data or the methods.
In this paper we investigate the possible contribution to this scatter
by nontrivial properties of the biasing scheme --- the deviations from 
linear biasing and the stochastic nature of the biasing scheme.
This is done using a simple and natural formalism for general galaxy biasing.

The biasing of density peaks in a Gaussian random field is well
formulated (\eg, 
Kaiser 1984; Bardeen \etal 1986)
and it provides a very crude theoretical framework for the origin of
galaxy density biasing.
In this scheme, the galaxy--galaxy and mass--mass correlation functions
are related in the linear regime via
\be
\xigg(r) =b^2 \ximm(r),
\label{eq:xi}
\ee
where the biasing parameter $b$ is a constant independent of scale $r$.
However, a much more specific linear biasing model is often assumed in common
applications, in which the local density fluctuation fields of galaxies and 
mass are assumed to be deterministically related via the relation
\be
\delg(\vx) = b\, \del(\vx).
\label{eq:linear}
\ee
Note that \eq{xi} follows from \eq{linear}, but the reverse is not true.

The deterministic linear biasing model is not a viable model.
It is based on no theoretical motivation.
If $b>1$, it must break down in deep voids because values of $\delg$
below $-1$ are forbidden by definition. 
Even in the simple case of no evolution in comoving galaxy number density,
the linear biasing relation is not preserved during the course
of fluctuation growth.
Non-linear biasing, where $b$ varies with $\del$, is inevitable.

Indeed, the theoretical analysis of the biasing of collapsed halos 
versus the underlying mass (Mo \& White 1996), using 
the extended Press-Schechter approximation (Bond \etal 1991),
predicts that the biasing is nonlinear and provides a useful approximation 
for its behavior as a function of scale, time and mass threshold.
$N$-body simulations provide a more accurate description
of the nonlinearity of halo biasing (see Figure~\ref{fig:1};
Somerville \etal 1998), and show that 
the model of Mo \& White is a good approximation.  
We provide more details about theoretical, numerical and observational
constraints on the exact shape of nonlinear biasing in 
\S~\ref{sec:constraints}, where we 
estimate the magnitude of nonlinear biasing effects.

It is important to realize that once the biasing is nonlinear at one 
smoothing scale, the smoothing
operation acting on the density fields guarantees that the biasing
at any other smoothing scale obeys a different functional form of
$b(\del)$ and is also non-deterministic.
Thus, any deviation from the simplified linear biasing model
must also involve both scale-dependence and scatter.

The focus of this paper is therefore on the consequences of the 
stochastic properties of the biasing process, which could either be
related to the nonlinearity as mentioned above, or arise from other
sources of scatter.  An obvious part of this stochasticity
can be attributed to the discrete sampling of the density field by galaxies
--- the shot noise. 
In addition, a statistical, physical scatter in the efficiency of galaxy 
formation 
as a function of $\del$ is inevitable in any realistic scenario.
It is hard to believe that the sole property affecting the efficiency of
galaxy formation is the underlying mass density at a certain smoothing scale
(larger than the scale of galaxies). For example, the random variations in the
density on smaller scales is likely to be reflected in the efficiency 
of galaxy formation. As another example,
the local geometry of the background structure, via the deformation 
tensor, must play a role too. In this case, the three eigenvalues
of the deformation tensor are relevant parameters.  Such hidden 
variables would show up as physical scatter in the density-density relation.
A similar scatter is noticeable in the distribution of particular 
morphological types versus the underlying total galaxy distribution
(Lahav \& Saslaw 1992).
The hidden scatter is clearly seen for halos in simulations including 
gravity alone (\S~\ref{sec:constraints} below and Figure~\ref{fig:1}
based on Somerville \etal (1998)
even before the more complex processes involving gas dynamics, star formation
and feedback affect the biasing and in particular its scatter.

In practice, there are two alternative options for dealing with the 
shot noise component of the scatter.
In some cases one can estimate the shot noise and try to remove it
prior to the analysis of measuring $\beta$. This is sometimes
difficult, \eg,  because of the finite extent of the galaxies which introduces
anti-correlations on small scales. The shot noise is especially large
and hard to estimate in the case where the biasing refers to light 
density rather than number density.
The alternative is to treat the shot noise as an intrinsic part of the local 
stochasticity of the biasing relation without trying to separate it from
the physical scatter. The formalism developed below is valid in either
case.

In \S~\ref{sec:local} we present the biasing formalism, separate the
effects of nonlinear biasing and stochastic biasing, and apply the formalism
to measurements involving local second-order moments of $\del$ and $\delg$.
In \S~\ref{sec:cor} we derive relations for two-point correlation functions
in the presence of local biasing scatter.
In \S~\ref{sec:dist} we apply the formalism to the analysis of redshift
distortions.
In \S~\ref{sec:high} we address methods involving third-order moments.
In \S~\ref{sec:constraints} we discuss constraints on the 
nonlinearity and scatter in the biasing scheme based on simulations,
simple models and observations. 
In \S~\ref{sec:conc} we summarize our conclusions and discuss our results
and future prospects.

\section{LOCAL MOMENTS}
\label{sec:local}

\subsection{Conditional Distribution}
\label{sec:local_def}


Let $\del(\vx)$ ($\equiv \delta \rho/\rho$) be the field of mass-density 
fluctuations and $\delg(\vx)$ ($\equiv \delta n/n$)
the corresponding field of galaxy-density 
fluctuations (or, alternatively, the field of light-density fluctuations, 
whose time evolution is less sensitive to galaxy mergers). 
The fields are both smoothed with a fixed smoothing window, which
defines the term ``local". The concept of galaxy biasing is meaningful
only for smoothing scales larger than the comoving
scale of individual galaxies, namely
a few Mpc. An example would thus be a top-hat window of radius $8\hmpc$,
for which the rms fluctuations of optical galaxies is about unity. 
Our analysis is confined to a specific smoothing scale,
at a specific time, and for a specific type of objects.             

Assume that both $\del$ and $g$
are random fields, with one-point probability distribution 
functions (PDF) $P(\del)$ and $P(g)$, both of zero mean by definition
and of standard deviations $\sig^2\equiv\av{\del^2}$ and
$\sg^2\equiv\av{g^2}$.
The key idea is to consider the local biasing relation between galaxies and
mass to also be a {\it random\,} process, specified by the 
{\it biasing conditional distribution\,} $P(g |\del)$ of $g$ at a given $\del$.

\def\fa{p}
\def\fb{q}

We shall use below in several different ways the following lemma 
relating joint averaging and conditional averaging
for any functions $\fa(g)$ and $\fb(\del)$: 
\be
\av{\fa(g) \fb(\del)} = \av{\,\av{\fa(g)|\del}_{g|\del}\,\fb(\del)\, }_\del,
\label{eq:lemma}
\ee
where the inner average is over the conditional distribution
of $g$ at a given $\del$ and the outer average is over the
distribution of $\del$.
This is true because
\be
\av{\fa(g) \fb(\del)} = \int\int dg\,d\del\, P(g,\del)\, \fa(g) \fb(\del)
= \int d\del\, P(\del)\, \fb(\del)\, \int dg\, P(g|\del)\, \fa(g) ,
\ee
in which the first equality is by definition,
and where for the second equality $P(g,\del)$ has been replaced by
$P(g|\del) P(\del)$ and the double integration has become successive.

\subsubsection{Conditional Mean -- Nonlinearity}

Define the {\it mean biasing function\,} $b(\del)$ by the conditional mean,
\be
b(\del)\,\del\equiv\av{g|\del} 
=  \int d g \, P(\delg\vert\del)\, g .
\ee
This function is plotted in Figure~\ref{fig:1}.
This is a natural generalization of the deterministic linear biasing relation, 
$g=b_1\del$.
The function $b(\del)$ allows for any possible {\it non-linear\,} biasing and
fully characterizes it; it reduces to the special case of {\it linear\,} 
biasing when $b(\del)=b_1$ is a constant independent of $\del$. 

In the following treatment of second-order local moments,
we will find it natural to characterize the function $b(\del)$
by the moments $\bh$ and $\bt$ defined by
\be
\bh \equiv\ \av{b(\del)\, \del^2} /\sig^2
\quad {\rm and} \quad
\bt^2 \equiv\ \av{b^2(\del)\, \del^2} /\sig^2 . 
\ee
In the case of linear biasing they both coincide with $b_1$.  
It will be shown that the parameter $\bh$ is the natural generalization 
of $b_1$
and that the ratio $\bt/\bh$ is the relevant measure of nonlinearity in the 
biasing relation;
it is unity for linear biasing, and is either larger or smaller than unity
for nonlinear biasing. 
As can be seen from the definitions of 
$\bh$ and $\bt$, this measure is independent of the stochasticity 
of the biasing.  It thus allows one to maintain general nonlinearity
while addressing stochasticity.

\subsubsection{Conditional Variance -- Stochasticity}

The local {\it statistical\,} 
character of the biasing relation can be expressed 
by the conditional moments of higher order about the mean at a given $\delta$. 
Define the {\it random biasing field\,} $\eps$ by
\be
\eps \equiv g -\av{g |\del} ,
\label{eq:eps}
\ee
such that its local conditional mean vanishes by definition, 
$\la \eps | \del \ra =0$.
The local variance of $\eps$ at a given $\del$ defines the 
{\it biasing scatter function\,} $\sb(\del)$ by
\be
\sb^2(\del) \equiv \av{\eps^2 | \del}/\sig^2.
\label{eq:sbd}
\ee
The scaling by $\sig^2$ is for convenience.
The function $\av{\eps^2 | \del}^{1/2}$ is marked by error bars in
Figure~\ref{fig:1}.
By averaging over $\del$, and using \eq{lemma},
one obtains the constant of local {\it biasing scatter\,},
\be
\sb^2 \equiv \av{\eps^2}/\sig^2 .
\label{eq:sb}
\ee

Thus, to second order, the nonlinear and stochastic biasing relation
is characterized locally by three basic parameters: $\bh$, $\bt$, and $\sb$.
The parameters $\bh$ and $\bt/\bh$ refer to the mean biasing and its 
nonlinearity, while $\sb/\bh$ measures the scatter. 
This parameterization separates
in a natural way the properties of nonlinearity and stochasticity.
The formalism simply reduces to the case of linear biasing when $\bh=\bt$
and to deterministic biasing  when $\sb=0$.

If the biasing conditional distribution, $P(\eps|\del)$, is a Gaussian 
[still allowing $b(\del)$ and $\sb^2(\del)$ to vary with $\del$], 
then the first- and
second-order moments fully characterize the biasing relation.
In much of the following we will restrict ourselves to second moments,
but we shall see in \S~\ref{sec:high} that a generalization to higher 
biasing moments, such as the skewness, is straightforward.

\subsection{Variances and Linear Regression}
\label{sec:local_var}


From the basic parameters defined above one can derive other useful
biasing parameters. 
A common one is the ratio of {\it variances\,},
sometimes referred to as ``the" biasing parameter, 
\be
\bv^2 \equiv {\sg^2 \over \sig^2} 
= \bt^2 +\sb^2 .
\label{eq:bv}
\ee
The second equality is an interesting result of \eq{lemma}. 
It immediately shows that $\bv$ is sensitive both to nonlinearity and
to stochasticity, and that always $\bv \geq \bt$.
Note the roles of $\bt^2$ and $\sb^2$ as the respective contributions 
of biasing nonlinearity and biasing scatter to the total scatter in $g$.
This makes $\bv$ biased compared to $\bh$,  
\be
\bv= \bh\, \left( {\bt^2\over \bh^2} +{\sb^2\over \bh^2} \right)^{1/2} ,
\label{eq:bv1}
\ee
by the root of the sum in quadrature of 
the nonlinearity factor $\bt/\bh$ and the scatter factor $\sb/\bh$.

Using \eq{lemma}, the mean parameter $\bh$ is simply related to the
{\it covariance\,}, 
\be
\bh\sig^2 = \av{g \del} .
\label{eq:gm}
\ee
Thus, $\bh$ is the slope of the linear regression of $g$ on $\del$
and it serves as the basic biasing parameter --- the natural generalization
 of the linear biasing parameter $b_1$.
Unlike the variance $\av{g^2}$ in \eq{bv}, the covariance in \eq{gm}
has no additional contribution from the biasing scatter $\sb$.

A complementary parameter to $\bv$ is the 
{\it linear correlation coefficient\,},
\be
\cor \equiv {\av{g \del} \over \sg \sig}
  = {\bh \over \bv} 
  =
\left( {\bt^2\over \bh^2} +{\sb^2\over \bh^2} \right)^{-1/2} .
\label{eq:r}
\ee
The equalities are based on \eq{gm} and \eq{bv1} respectively.

The ``inverse" regression, of $\del$ on $g$, yields another biasing parameter:
\be
\binv\equiv{\sg^2 \over \av{g\del}} 
  ={\bv \over \cor}  
  ={\bh \over \cor^2}
  ={\bv^2 \over \bh}
  =\bh \left( {\bt^2\over\bh^2} + {\sb^2\over\bh^2} \right) .
\label{eq:binv}
\ee
The parameter $\binv$ is closer to what is measured in reality 
by two-dimensional linear regression
(\eg, Sigad \etal 1998), 
because the errors in $\del$ 
are typically larger than in $g$.   
Note that $\binv$ is biased compared to $\bh$, with the ratio
given by the sum in quadrature of the nonlinearity factor $\bt/\bh$ and the 
scatter factor $\sb/\bh$.


It is worthwhile to
summarize the relations between the parameters in the two degenerate cases.
In the case of {\it linear\,} and stochastic biasing, 
the above parameters reduce to
\be
\bt=\bh=b_1,\quad
\bv=b_1 \left( 1+{\sb^2\over b_1^2} \right)^{1/2} ,\quad
\cor={b_1 \over \bv} ,\quad
\binv=b_1 \left( 1+{\sb^2\over b_1^2} \right) .
\label{eq:lin-sto}
\ee
Thus, $b_1 \leq \bv \leq \binv$.
 
In the case of nonlinear and {\it deterministic\,} biasing they reduce 
instead to
\be
\bt\neq\bh,\quad
\sb=0,\quad
\bv=\bt,\quad
\cor={\bh\over\bt},\quad
\binv={\bt^2\over\bh} .
\label{eq:nl-det}
\ee
Both $\binv$ and $\bv$ are biased with respect to $\bh$,
and the bias in $\binv$ is always larger.
Whether they are biased high or low compared to $\bh$
depends on whether the nonlinearity factor $\bt/\bh$ is larger or smaller 
than unity, respectively. 
We shall see in \S~\ref{sec:constraints} that 
although $\bh$ and $\bt$ could significantly
differ from unity and from $b(\del=0)$, the ratio $\bt/\bh$, in realistic 
circumstances, typically obtains values in the range 
$1.0 \leq \bt/\bh \leq 1.1$.  This means that the effects of nonlinearity 
are likely to be relatively small.
 
In the fully degenerate case of linear and deterministic biasing,
all the $b$ parameters are the same, and only then $\cor=1$.
Note, again, that the parameters $\bh/\bt$
and $\sb/\bt$ nicely separate the 
properties of nonlinearity and stochasticity, while the parameters $\bv$, $r$
and $\binv$ mix these properties.


In actual applications,
the above local biasing parameters are involved when the parameter ``$\beta$"
is measured from observational data in several different ways.
For linear and deterministic biasing this parameter is defined unambiguously
as $\beta_1\equiv f(\Omega)/b_1$. But any deviation from this
degenerate model causes us to actually measure different $\beta$ parameters 
by the different methods.

For example,
it is the parameter $\betav \equiv f(\Omega)/\bv$ which is 
determined from measurements of $\sg$ and $\sig f(\Omega)$.
The former is typically determined from a redshift survey,
and the latter either from an analysis of peculiar velocity data,
or from the abundance of rich clusters (with a slightly modified $\Omega$
dependence), 
or by COBE normalization of a specific power-spectrum shape
for mass density fluctuations. 
As noted in \eq{bv}, in the case of stochastic biasing $\bv$
is always an overestimate of $\bt$. When the biasing is linear, \eq{lin-sto},
$\bv$ is an overestimate of $b_1$.
The corresponding $\betav$ is thus underestimated accordingly.

A useful way of estimating $\beta$ (\eg, Dekel \etal 1993; Hudson \etal
1995; Sigad \etal 1998) is via
the {\it linear regression\,} of the fields in our cosmological 
neighborhood, \eg, $-\divv(\vx)$ on $\delg(\vx)$
(or, alternatively, via a regression of the corresponding velocities).
In the mildly-nonlinear regime, $-\divv(\vx)$ is actually replaced by
another function of the peculiar velocity field and its first spatial
derivatives, which better approximates the scaled mass-density field
$f(\Omega)\del(\vx)$ (\eg, Nusser \etal 1991).
The regression that is done, taking the errors on both sides into account,
is effectively $\del$ on $\delg$, because the errors in
$\divv$ (or $f\del$) are typically more than twice as large as the
errors in $\delg$ (\eg, Sigad \etal 1998).
Hence, the parameter that is being measured is close to
$\beta_{\rm inv}\equiv f(\Omega)/\binv$.
In the case of linear gravitational instability
and linear deterministic biasing,
the slope of this regression line is simply $\beta_1$.
Note in \eq{lin-sto} that in the case of linear and stochastic biasing $\binv$
is an overestimate of $b_1$.
The corresponding $\beta$ is thus underestimated accordingly
in this inverse-regression analysis.

\section{TWO-POINT CORRELATIONS}
\label{sec:cor}

For the purpose of redshift-distortion analysis we need to generalize
our treatment of stochastic and nonlinear biasing to deal with spatial
correlations.
Given the random biasing field $\eps$, \eq{eps}, 
we define the two-point biasing--matter cross-correlation function
and the biasing auto-correlation function by
\be
\xibm(r) \equiv \av{\eps_1 \del_2} ,
\quad
\xibb(r) \equiv \av{\eps_1 \eps_2} , 
\label{eq:xibb}
\ee
where the averaging is over the ensembles at points 1 and 2 separated by $r$.
(Recall that throughout this paper the fields  
are assumed to be smoothed with a given window.)
By the definition of the random biasing field $\eps$,  
and the local scatter, \eq{sb}, at zero lag one has
$\xibm(0)=0$ and $\xibb(0)=\sb^2\sig^2$.
We now define the biasing as {\it local\,} if
\be
\xibm(r)=0 \ \ {\rm for\ any\ } r, 
\quad {\rm and} \quad
\xibb(r)=0 \ \ {\rm for\ } r>\xb ,
\ee
where $\xb$ is typically on the order of the original smoothing scale.
(Some implications of local biasing are discussed in   
Scherrer \& Weinberg 1998.)                            

A two-point equivalent lemma to \eq{lemma} implies that
\be
\la g_1 \dtwo \ra = \la \la g_1 \dtwo \,| \done \dtwo \ra_{g|\del} \ra_\del 
\quad{\rm and}\quad
\la g_1 g_2 \ra = \la \la g_1 g_2 \,| \done \dtwo \ra_{g|\del} \ra_\del .
\label{eq:lemma4}
\ee
Using these identities, one obtains analogous relations to \eqs{gm} 
and \eqss{bv}:
\be
\xigm(r) \equiv \av{g_1 \del_2}
= \av{b(\done)\done\, \dtwo} + \xibm(r) ,
\label{eq:xigm}  
\ee
\be
\xigg(r) \equiv \av{g_1 g_2}
= \av{b(\done)\done\, b(\dtwo)\dtwo} + \xibb(r) .
\label{eq:xigg} 
\ee

In the case of {\it linear\,} and {\it local\,} biasing, 
these become
\be
\xigm(r)=b_1 \ximm(r) ,
\label{eq:xigm_lin}
\ee
\be
\xigg(r) = b_1^2 \ximm(r) +\xibb(r) ,
\quad {\rm with\ } \xibb(r)=0 \ \ {\rm for\ } r>\xb .
\label{eq:xigg_lin}
\ee
Note that the biasing parameter that appears here is $b_1$, not to be confused
with $\bv$ when the biasing is stochastic.

To see how the power spectra are affected by the biasing scatter,
we assume, without limiting the generality of the analysis, 
that the local biasing can be approximated by a step function:  
\be
\xibb(r)=\cases {
                 \sb^2\sig^2   & $r<\xb$ \cr
                 0             & $r>\xb$       } .
\label{eq:xibb2}
\ee
Recalling that the power spectra are the Fourier transforms of
the corresponding correlation functions,
\be
P(k) = 4\pi \int_0^\infty \xi(r) {\sin (kr) \over kr} r^2 dr ,
\label{eq:P-xi}
\ee
we get for $k\ll\xb^{-1}$, from \eqss{xigm_lin} and \eqss{xigg_lin},
\be
\Pgm(k)= b_1 \Pmm(k) ,
\label{eq:Pgm}
\ee
\be
\Pgg(k)= b_1^2\Pmm(k) + \sb^2\sig^2 V_{\rm b} ,
\label{eq:Padd}
\ee
where $V_{\rm b}$ is the volume associated with the original smoothing
length, $\xb$.
We see that the {\it local\,} biasing scatter adds to $\Pgg(k)$ an
{\it additive constant\,} at all $k$ values.

Finally, one can address the effect of the scatter on moments of
the fields smoothed at a general smoothing length $r>\xb$.
Each of these moments is related to the corresponding power spectrum via
an integral of the form
\be
\la \del \del \ra_r = \int d^3k\, \tilde W^2(kr)\, P(k),
\label{eq:smooth}
\ee
where $\tilde W(kr)$ is the Fourier Transform of the smoothing window
of radius $r$.
Using \eq{Padd} one obtains for linear biasing
\be
\sg^2(r)=b_1^2 \sig^2(r) +\sb^2\sig^2 (\xb/r)^3 .
\label{eq:sadd}
\ee
In the following section we will apply this formalism to the linear analysis 
of redshift distortions.

For the purpose of an analysis involving {\it nonlinear\,} biasing,
note that two-point averages as in \eqs{xigm} and \eqss{xigg} 
are calculable once one knows the function $b(\del)$ and the 
one- and two-point distributions of the underlying density field $\del$.
It turns out that the relation for $\xigm$, \eq{xigm}, is in fact a 
simple extension of \eq{gm} involving only the local moment $\bh$:
\be
\xigm(r)=\bh \ximm(r) .
\label{eq:xigm_bh}
\ee
To prove this, we use $P(\done,\dtwo)=P(\dtwo|\done)P(\done)$ to write
\be
\av{b(\done)\done\, \dtwo} =
\int d\done P(\done) b(\done)\done\, \int d\dtwo P(\dtwo|\done)\dtwo ,
\ee
and use the fact that $\av{\dtwo |\done}= \done \ximm(r)/\sig^2$.
To compute $\bh$ one needs to know only the function $b(\del)$ and 
the one-point distribution $P(\del)$.
Higher-order moments, like the one in \eq{xigg}, would, in general,
involve the two-point PDF as well.

\section{REDSHIFT DISTORTIONS}
\label{sec:dist}

A very promising way of estimating $\beta$ is via redshift distortions
in a redshift survey
(\eg, Kaiser 1987, Hamilton 1992, 1993, 1995, 1997;
Fisher \etal 1994; 
Fisher, Scharf \& Lahav 1994;  Heavens \& Taylor 1995;
Cole, Fisher \& Weinberg 1995; Fisher \& Nusser 1996; Lahav 1996).
Peculiar-velocity gradients along the line of sight distort  
the comoving volume elements in redshift space compared to the
corresponding volumes in real space. As a result, a large-scale isotropic 
distribution of galaxies in real space is observed as an anisotropic 
distribution in redshift space.  The relation between peculiar velocities and
mass density depends on $\Omega$, and hence the distortions relative 
to the galaxy density depend both on $\Omega$ and on the galaxy biasing
relation. 
In the deterministic and linear biasing case, this relation involves 
a single $\beta$ parameter (see Hamilton 1997). 
However, in the general biasing case, the distortion analysis is 
in principle complicated by the fact that the galaxies play two 
different roles; they serve both as luminous tracers of the mass 
distribution as well as test bodies for the peculiar-velocity field.

To first order, the local galaxy density fluctuations in redshift space 
($\delgs$) and real space ($\delg$) are related by
$\delgs = \delg - \pa u/\pa r$,
where $u$ is the radial component of the galaxy peculiar velocity $\vv(\vx)$.
Assuming no velocity biasing, linear GI theory predicts 
$\pa u/\pa r = - \mu^2 f(\Omega) \del$,
where $\mu^2$ is a geometrical
factor depending on the angle between $\vv$ and $\vx$.
Thus, the basic linear relation for redshift distortions is
\be
g_{\rm s}=g+f\mu^2\del .
\label{eq:dist_basic}
\ee


A general {\it local\,} expression for redshift distortions is obtained
by taking the mean square:
\be
\av{\delgs\delgs} = \av{g g} + 2(f\mu^2)\, \av{g \del}
                                 + (f\mu^2)^2\, \av{\del\del} .
\label{eq:dist_loc}
\ee
With our formalism for stochastic biasing, using \eq{bv} and \eq{gm},
it becomes
\be
\sgs^2 = \sg^2 [1 + 2(f\mu^2) \cor \bv^{-1}  + (f\mu^2)^2 \bv^{-2}].
\label{eq:dist_loc_b}
\ee
This is similar to equation (7) of Pen (1998), in the sense that it involves
both $\bv$ and $\cor$ in a non-trivial way
and is thus directly affected by the stochasticity of the biasing scheme. 
However, we shall see that
this is true {\it only\,} for the {\it local\,} moments, 
at the original smoothing
length for which $\bv$ and $\cor$ were defined. 
When the biasing is stochastic, the situation at non-zero lag is very
different.

\subsection{General Linear Redshift Distortions at Non-zero Lag}

The general linear analysis of redshift distortions involves non-local
analysis.  The general expression in terms of
correlation functions 
is obtained straightforwardly from \eq{dist_basic} by averaging 
$\av{g^s_1 g^s_2}$ over the distributions of $\del$ at a pair of 
points separated by $r$: 
\be
\xigg^s(r) = \xigg(r) + 2(f\mu^2)\, \xigm(r) + (f\mu^2)^2\, \ximm(r) .
\label{eq:dist_xi}
\ee
(Recall that our correlation functions and power spectra
correspond to the smoothed fields.)

Recalling that the power spectra are the Fourier transforms of
the corresponding correlation functions, \eq{P-xi},
one can equivalently write
\be
\Pgg^s(k) = \Pgg(k) + 2(f\mu^2)\, \Pgm(k) + (f\mu^2)^2\, \Pmm(k) .
\label{eq:dist_P}
\ee
[This expression can alternatively be obtained from the fact that
\eqs{dist_loc} and \eqss{smooth} are valid for any smoothing scale $r$.]

Similarly, the spherical harmonic analysis for redshift distortions
in linear theory for a flux-limited survey
(Fisher, Scharf \& Lahav 1994, equation 11)
can be extended to yield for the mean-square harmonics in redshift
space:
\be
\la | a_{lm}^s |^2 \ra =
{2\over\pi} \int dk\,k^2\, [
    | \Psi_l (k) |^2\, P_{\rm gg}(k) \,
+\, 2 f | \Psi_l (k) \Psi_l^c(k) | \, P_{\rm gm} (k) \,
+\, f^2 | \Psi_l^c (k) |^2\, P_{\rm mm}(k) ] ,
\ee
where $\Psi_l(k)$ is the real-space window function
and $\Psi_l^c(k)$ is the redshift-correction window function,
both depending on the selection function and the weighting function.
Again, this expression mixes the three different power spectra
such that the $\Omega$ dependence, in general, may involve more than one 
unique $\beta$.

The crucial question is how to relate the correlation functions, or
power spectra, to the biasing scheme.
In the case of linear and deterministic biasing, one simply has
$\Pgg=b_1\Pgm=b_1^2\Pmm$, so the distortion relation reduces to Kaiser's
formula,
\be
\Pgg^s=\Pgg(1+\mu^2\beta_1)^2 ,
\ee
where $\beta_1\equiv f(\Omega)/b_1$ (and here $b_1=\bv$).
In order to obtain more specific distortion relations for the case of 
stochastic biasing, we shall use the spatial correlations from 
\S~\ref{sec:cor} in the general distortion relations of the current section.

\subsection{Distortions for Linear, Stochastic and Local Biasing}

At zero lag, by definition, $\xibb(0)=\sb^2\sig^2$. Then, as in \eq{bv}
for the local moments, $\xigg(0)=\bv^2\ximm(0)$,
and the general distortion relation for $\xi$, \eq{dist_xi}, reduces to
an equation similar to the local \eq{dist_loc_b}, 
in which the second and third terms involve different combinations  
of $\cor$ and $\bv$ and thus allow to determine them separately.    
However, at large separations $r>\xb$, where $\xibb$ vanishes by the
assumption of local biasing, one obtains instead,
from \eqs{xigm_lin} and \eqss{xigg_lin}
\be
\xigg^s(r) = \xigg(r) [1 + 2(f\mu^2)\, b_1^{-1} + (f\mu^2)^2\, b_1^{-2}] .
\label{eq:dist_xi_lin}
\ee
This is the degenerate Kaiser formula, which is very different from the 
expression for local moments, \eq{dist_loc_b}. 
In particular, it is independent of the biasing scatter! 
It now involves only the mean biasing parameter $b_1$ (a
degenerate combination of $\cor$ and $\bv$), but it
contains no information on the stochasticity, $\sb$.
In terms of $b_1$, the relation for $\xi$ is identical to
the case of deterministic biasing.
Thus, the distortion analysis at $r>\xb$ 
is indeed incapable of evaluating the stochasticity of the process.
This is a straightforward result of the assumed locality of the biasing scheme.
The biasing scatter at two distant points is uncorrelated and therefore its
contribution to $\xigg$ cancels out.

On the other hand, the redshift distortion analysis is sensitive to 
the {\it non-linear\,} properties of the biasing relation. A proper analysis
would require a nonlinear treatment of the redshift distortions including
a nonlinear generalization of the GI relation $\divv = -f\del$,
because the nonlinear effects of biasing and gravity enter at the same
order.
The result is more complicated than \eq{dist_xi}, but is calculable
in principle once one knows the function $b(\del)$ and the one-
and two-point probability distribution functions of $\del$.
The insensitivity to stochasticity remains valid in the case of
nonlinear biasing.

Back to the case of linear stochastic biasing.
The distortion relation for $P(k)$ becomes more complicated because
of the additive term in \eq{Padd}.  For linear biasing, when substituting 
\eq{Padd} in \eq{dist_P}, the terms analogous to the ones
involving $b_1^{-1}$ and $b_1^{-2}$ in \eq{dist_xi_lin} for $\xi$
are multiplied by $[1-\sb^2\sig^2 V_{\rm b}/\Pgg(k)]$, a function of $k$.  
The distortion relation for $P(k)$ is thus affected by the biasing 
scatter in a complicated way.

However, if the scatter is small, there may be a $k$ range around the peak
of $P(k)$ where the additive scatter term in \eq{Padd} is small compared
to the rest. In this range the relation reduces to an expression
similar to \eq{dist_xi_lin} for the corresponding power spectra.
For example, 
if $\xb \sim 8\hmpc$ we have $V_{\rm b} \sim 2\times 10^3 (\hmpc)^3$,
while $\Pmm(k)\sim 10^4 (\hmpc)^3$ at the peak (\eg, Kolatt \& Dekel 1997), 
so a significant $k$ range
of this sort is viable, especially if $\sb\sig \ll b_1$.
On the other hand, the scatter term always dominates \eq{Padd} at small
and at large $k$'s. 
If $\sb\sig \sim 1$, then the scatter may dominate already not much
below $k \sim 0.01 (\hmpc)^{-1}$.

In terms of moments of a general smoothing length $r>\xb$,
using \eq{sadd} in \eq{dist_loc} one obtains
a complicated distortion relation again.
For small scatter there may be a limited range of scales for which the
first term in \eq{sadd} dominates and then the distortions reduce to an 
equation similar to \eq{dist_xi_lin} for moments of smoothed fields.
At large enough scales, where $\Pmm$ is rising with
$k$ and thus $\sig^2$ is decreasing faster than $\propto x^{-3}$,
the scatter becomes dominant.

Note that in order to obtain equation (7) of Pen (1998) from the 
general linear distortion relation, \eq{dist_P}, one has to define 
$k$-dependent biasing parameters by $\Pgg(k)=\bv(k)^2\Pmm(k)$
and $\Pgm(k)=\bv(k)\cor(k) \Pmm(k)$.
(Note that Pen's $\beta$ refers to his $b_1$, which is equivalent to 
our $\bv$, except that he allows it to vary with $k$).
In the case of local biasing, a comparison to our \eq{Pgm} and \eq{Padd}
yields $\bv(k)^2 = b_1^2 + \sb^2\sig^2 V_{\rm b}/\Pmm(k)$
and $\bv(k) \cor(k) = b_1$.
In the $k$ range near the peak of $\Pmm(k)$ where the constant term
in \eq{Padd} may be negligible, one has $\bv(k) = b_1$ and $\cor(k)=1$,
and there is indeed no sign of the stochasticity in the distortion relation.

\section{SKEWNESS AND THREE-POINT CORRELATIONS}
\label{sec:high}

\subsection{Skewness}
\label{sec:high_skew}

We now move to measures of biasing involving third-order moments.
Given the biasing random fields $\eps$, define the biasing skewness function
$\Sb(\del)$, in analogy to $\sb(\del)$ of \eq{sbd}, by
\be
\Sb(\del) S \equiv \av{ \eps^3 | \del} ,
\label{eq:skewb}
\ee
where $S\equiv \av{\del^3}$. After averaging over $\del$, the biasing skewness
parameter is
\be
\Sb \equiv \av {\eps^3}/S .
\ee
The biasing parameter that is defined by ratio of skewness moments is then,
based on \eq{lemma} and after some algebra, 
\be
b_3^3\equiv {\av{g^3} \over \av{\del^3} } =
    {\av{\del^3 b^3(\del) } \over S} 
    + {3\sig^2 \av{\del b(\del) \sb^2(\del)} \over S} 
    +\Sb .
\ee

In the case of deterministic biasing, $\Sb=\sb=0$, one has
\be
b_3^3 = {\av{\del^3 b^3(\del)} \over S} ,
\ee
which differs from the parameters $\bh$ and $\bt$ of \S~\ref{sec:local} 
due to nonlinear effects. 
In the linear case where both $b(\del)$ and $\sb(\del)$ are constants,
the expression for $b_3$ reduces to
\be
b_3 = b_1 \left( 1+ {\Sb \over b_1^3} \right)^{1/3} .
\ee
Now, if $\Sb=0$, then $b_3=b_1$ independently of $\sb$.
If, on the other hand, $P(g|\del)$ is positively skewed, then $b_3>b_1$.

An interesting quantity involving the skewness and variance of $\del$
is $S_3\equiv S/\sigma^4$. In the second-order approximation to
GI with Gaussian initial fluctuations this quantity is constant in time.
For top-hat smoothing and a given power spectrum
of an effective power index $n$ at the smoothing scale,
this constant is $S_3=34/7-(3+n)$
(Juszkiewicz \etal 1993).
The corresponding quantity involving the moments of $\delg$,
$S_{\rm 3g}\equiv {\la\delg^3\ra / \la\delg^2\ra^2}$,
provides an observational measure of biasing (Weinberg 1995):
\be
b_{\rm S3}^{-1} \equiv {S_{\rm 3g} \over S_3}
={b_3^3 \over \bv^4}                                 
={ \av{\del^3 b^3(\del)}/S  +3\sig^3 \av{\del b(\del) \sb^2(\del)}/S +\Sb
 \over
 [\av{\del^2 b^2(\del)}/\sig^2 +\sb^2]^2 }.
\ee

In the case of deterministic biasing,
\be
b_{\rm S3} = { \av{\del^2 b^2(\del)}^2 /\sig^4
              \over
               \av{\del^3 b^3(\del)} /S } .
\ee
In the case of linear biasing where $b(\del)$ and $\sb(\del)$ 
are constants, this ratio reduces to
\be
b_{\rm S3} = b_1\, {(1+\sb^2/b_1^2)^2 \over 1+\Sb/b_1^3}.
\ee
The biasing parameter obtained this way thus depends both on $\sb$ and $\Sb$.
If $\Sb=0$, as when $P(g|\del)$ is Gaussian,
then $b_{\rm S3}=b_1\, (1+\sb^2/b_1^2)^2$,
and the deviation from $b_1$ is even larger than that of $\binv$ or $\bv$,
\eq{lin-sto}.
For positive biasing skewness $\Sb$, the parameter $b_{\rm S3}$ may in fact
become smaller than $b_1$.

Szapudi (1998) has shown that, under certain simplifying assumptions,
the biasing parameters (taken as two coefficients in a Taylor expansion,
\eq{taylor}) can be determined using three-point statistics (cumulant 
correlators).  The assumptions made are that the biasing is local, 
deterministic and scale independent, and that redshift distortions 
are negligible.  This approach should be generalized to the more 
realistic case of stochastic biasing and to allow other nontrivial 
features in the biasing scheme.

\subsection{Cosmic Virial Theorem and Energy Equation}
\label{sec:high_cvt}

The Cosmic Energy Equation (CE) (Peebles 1980, \S 74; Peebles 1993, eq.  20.11; 
Davis, Miller \& White 1997)   
can be used to determine $\Omega$ by relating the observed dispersion
of galaxy peculiar velocities to a spatial integral over the galaxy--mass
cross-correlation function, $\xigm(r)$.
The observable is the galaxy--galaxy auto-correlation function
$\xigg(r)$, so the corresponding biasing parameter is
\be
b_{\rm ce} = \la \delg\delg \ra / \la \delg \del \ra . 
\ee
At zero lag,  $b_{\rm ce} = \binv$ of \eq{binv}.
At non-zero lag, $b_{\rm ce}$ can be derived from \eqs{xigm} and \eqss{xigg}.
For linear biasing, \eqs{xigm_lin} and \eqss{xigg_lin}, 
one obtains at non-zero lag
$b_{\rm ce} = b_1$, which is different from $\binv$ if the
biasing is stochastic, \eq{lin-sto}.

The estimation of $\Omega$ via the Cosmic Virial Theorem (CV),
as applied to galaxy surveys (Peebles 1980, \S 75; 
Bartlett \& Blanchard 1996),  
relates the observed dispersion of galaxy--galaxy peculiar velocities
to a spatial integral over the 3-point galaxy--galaxy--mass cross-correlation 
function, $\xi_{\rm ggm}$ (divided by $\xi_{\rm gg}$).
The observable is the 3-point galaxy correlation function $\xi_{\rm ggg}$, 
so the corresponding biasing parameter is
\be
b_{\rm cv} = \la \delg\delg\delg \ra / \la \delg\delg \del \ra .
\ee
At zero lag, using \eq{lemma},
\be
b_{\rm cv} = {  \av{ \del^3 b^3(\del)} 
                +3\sig^2 \av{\del b(\del) \sb^2(\del)}
                +\Sb S 
              \over
                \av{ \del^3 b^2(\del)} 
                +\sig^2 \av{\del \sb^2(\del) } }.
\ee

In the case of deterministic but nonlinear biasing, 
$b_{\rm cv} = \av{\del^3 b^3(\del)} / \av{\del^3 b^2(\del)}$,
which in general differs from any of the biasing parameters
discussed so far.
If $b(\del)$ and $\sb(\del)$ are constants and the biasing is stochastic, 
then, at zero lag,
$b_{\rm cv} = b_3^3 / b_1^2 = b_1\, ( 1+ {\Sb / b_1^3} )$.

If the analysis is done on scales smaller than the biasing coherence length
$\xb$, then the local expressions are relevant. Otherwise, one needs to
appeal to three-point spatial correlations.
   
Note that in the case of CV or CE (which are valid on small scales)
the measured quantity may be $\Omega/b$ or $\Omega/b^2$,   
depending on the application, rather than $\beta$ which
is typical in applications based on the linear approximation to 
gravitational instability.

\section{CONSTRAINTS ON THE BIASING RELATION}
\label{sec:constraints}

In the scheme outlined above, the local biasing process at given scale, time
and galaxy type,
is characterized by the conditional probability density function 
$P(\delg\vert\del)$. 
The conditional mean, or the function $b(\delta)$, contains the information
about the mean biasing (via the parameter $\bh$) and the nonlinear 
features (\eg, via $\bh/\bt$). 
The first additional quantity of interest in the case
of non-negligible scatter in the biasing relation can be 
the conditional standard deviation, the function $\sb(\delta)$,
and its variance over $\del$, $\sb^2$.
In order to evaluate the actual effects of nonlinear and stochastic
biasing on the various measurements of $\beta$, one should first try
to constrain these functions or evaluate these parameters
from simulations, theoretical approximations, and observations.

\subsection{Preliminary Results from Simulations}

In an ongoing study,
Somerville \etal (1998) are investigating the biasing in high-resolution 
$N$-body simulations of several cosmological scenarios, both for galactic
halos and for galaxies as identified using semi-analytic models.
Earlier results from simulations were obtained, \eg,                
by Cen \& Ostriker (1992) and in more detail by Mo \& White (1996). 
We refer here to an example of the preliminary results of Somerville \etal,
in the context of our biasing formalism.
As our test case we use a representative cosmological model: 
$\Omega=1$ with a $\tau$CDM power spectrum which roughly obeys
the constraints from large-scale structure.
The simulation mass resolution is $2\times 10^{10} \msun$ inside a box
of comoving side $85\hmpc$. The present epoch is identified with the
time when the rms mass fluctuation in a top-hat sphere of radius
$8\hmpc$ is $\sigma_8=0.6$.

\begin{figure}[t]
{\includegraphics{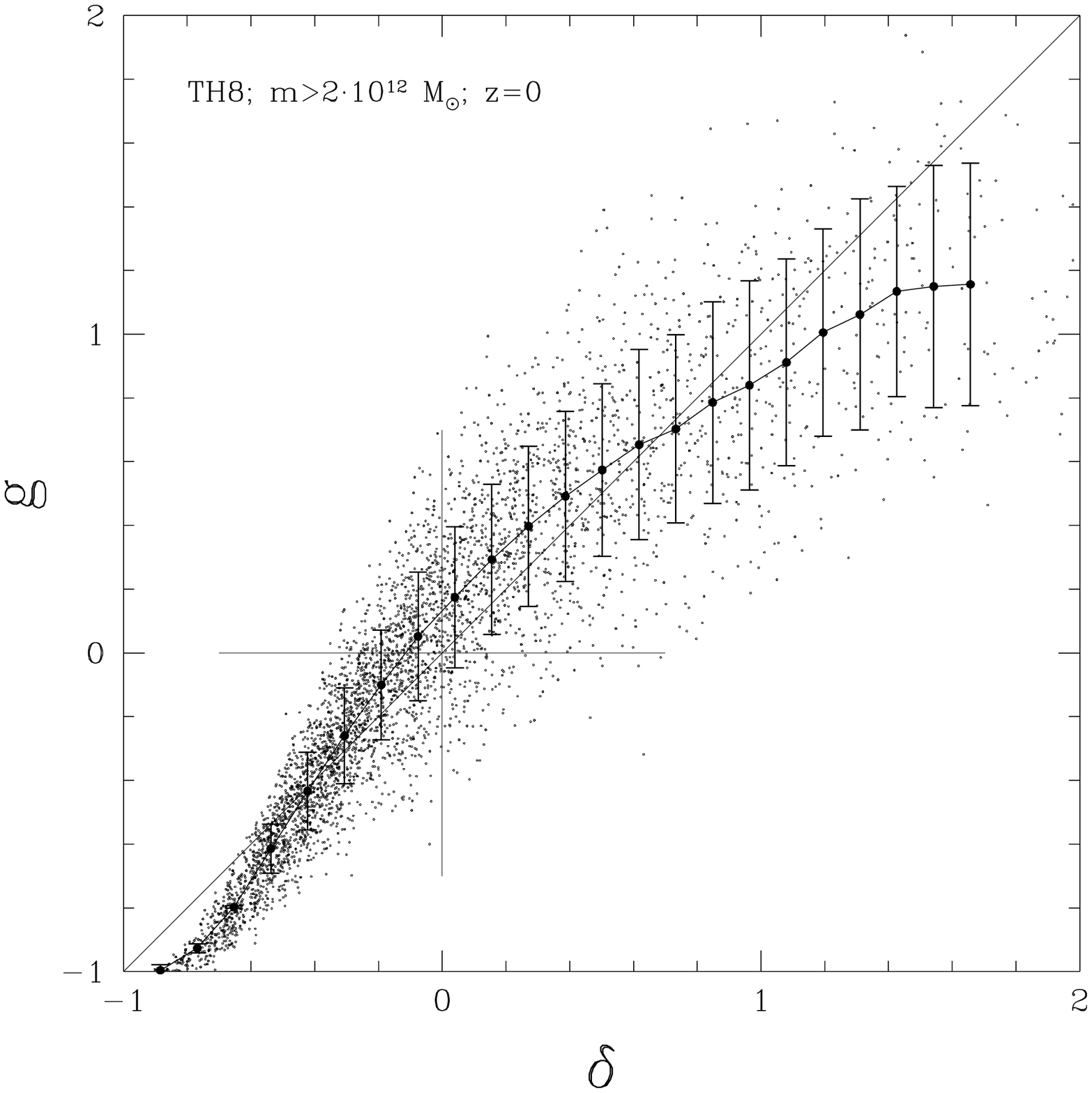}}
{\includegraphics{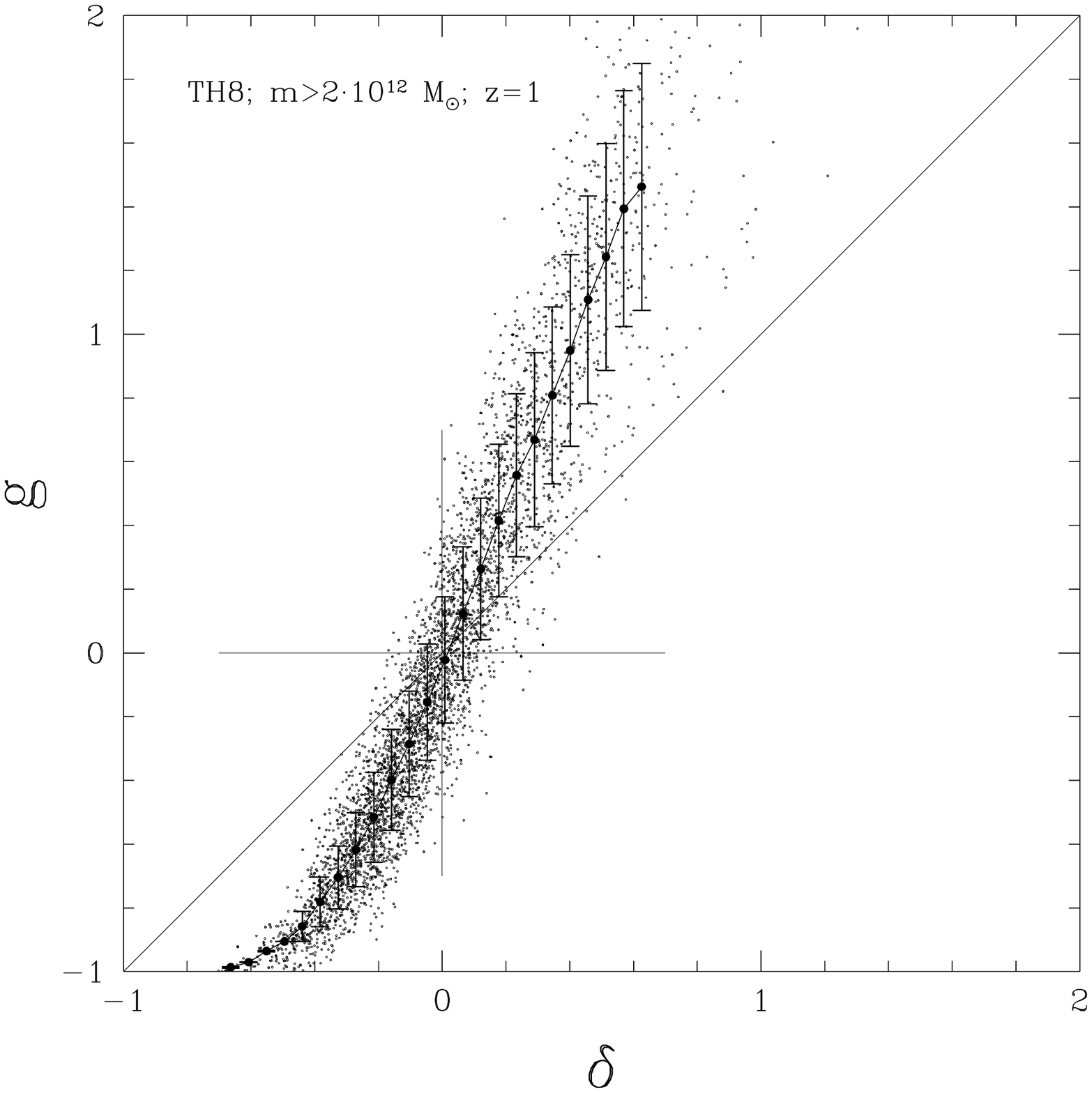}}
\vspace{9.0 cm}
\caption{\capt
Biasing of galactic halos versus mass in a cosmological $N$-body simulation,
demonstrating nonlinearity and stochasticity.
The conditional mean (solid curve) and scatter (error bars) are marked.
The fields smoothed with a top-hat window of radius $8\hmpc$
are plotted at the points of a uniform grid. 
Left: at the time when $\sigma_8=0.6$ (\eg, $z=0$).
Right: at an earlier time when $\sigma_8=0.3$ (\eg, $z=1$).
Based on Somerville \etal (1998).
}
\label{fig:1}
\end{figure}

Figure~\ref{fig:1} is borrowed from Somerville \etal (1998) in order to
demonstrate the qualitative features of the biasing scheme.
It shows the density fluctuation fields of galactic halos versus 
mass at the points of a uniform grid at two different times.  
The halos are selected above a mass threshold of $2 \times 10^{12} \msun$. 
The fields are smoothed with a top-hat window of radius $8\hmpc$.  
The conditional mean [$\av{g|\del}=b(\del)\del$] and the conditional scatter 
[$\av{\eps^2|\del}=\sb^2(\del) \sig^2$] are marked.

The nonlinear behavior in the negative regime, $\del <0$,
is characteristic of all masses, times, and smoothing scales:
the function $\av{g|\del}$ is flat near $g=\del=-1$ 
and it abruptly steepens towards $\del=0$.
In the positive regime, $\del >0$, the behavior is less robust --- 
it strongly depends on the mass, time and smoothing scale.
The scatter in the figure includes both shot noise and physical scatter
which are hard to separate properly.
The scatter function $\sb(\del)$ grows rapidly from zero at $\del=-1$ to 
a certain value near $\del=0$, and it continues to grow slowly
for $\del>0$ to an asymptotic value at large $\del$.
 
In the case shown at $z=0$,
the nonlinear parameter is $\bt^2/\bh^2=1.08$,
and the scatter parameter is $\sb^2/\bh^2=0.15$.
The effects of stochasticity and nonlinearity in this specific
case thus lead to moderate
differences in the various measures of $\beta$, on the order of $20-30\%$.
Gas-dynamics and other non-gravitational processes
may extend the range of estimates even further.

A recent hint for the origin of physical scatter in the biasing scheme
is provided by Blanton \etal (1998). They find, based on 
hydrodynamic cosmological simulations, that the local gas temperature,
which is an important factor affecting the efficiency of galaxy formation,
is not fully correlated with the other dominant factor --- the
local mass density. 
They therefore argue that this is a physical hidden variable that 
contributes to the stochasticity proposed in our current paper.  
They also detect significant scale dependence in the
biasing scheme, and identify its main source with the correlation 
of the temperature with the large scale gravitational potential.

\subsection{Approximations for Nonlinear Biasing}

Given the distribution $P(\del)$ of the matter fluctuations,
the biasing function $b(\del)$ should obey by definition at least the following 
two constrains: 
\begin{itemize}
\item
$g\geq -1$ everywhere because the galaxy density $\rho_{\rm g}$ 
cannot be negative,
and $g=-1$ at $\del=-1$ because there are no galaxies where there is 
no matter.
\item
$\av{g}=0$ because $g$ describes fluctuations about the mean galaxy density.
\end{itemize}

An example for a simple functional form that obeys the constraint at
$\delta=-1$ and reduces to the linear relation near $\delta=0$ is
(\eg, Dekel \etal 1993)
\be
\av{g|\del} = c\, (1+\del)^b -1 .
\label{eq:powerb}
\ee
The constraint $\av{g}=0$ is to be enforced by a specific choice of 
the factor $c$ as a function of the power $b$.
With $b>1$, this functional form indeed provides a reasonable fit 
to the simulated halo biasing relation in the $\delta < 0$ regime. 
However, the same value of $b$ does not 
necessarily fit the biasing relation in the $\delta > 0$ regime, 
which can require either $b<1$ or $b>1$ depending on halo mass, smoothing 
scale and redshift.
 
A better approximation could thus be provided by a combination of two
functions like \eq{powerb}
with two different biasing parameters $\bn$ and $\bp$ in the regimes
$\del\leq 0$ and $\del>0$ respectively. Another useful
version of such a combination is 
\be
\av{g|\del} = \cases {
 (1+b_0) (1+\del)^{\bn} -1 & $\delta \leq 0$ \cr
  \bp \del +b_0       & $\delta>0$ } ,
\ee
which provides an even better fit to the behavior in Fig.~\ref{fig:1}.
As mentioned above, the parameter $\bn$ is always larger than unity 
while $\bp$ ranges from slightly below unity to much above unity.
The best fit to Fig.~\ref{fig:1} at $z=0$ has $\bn \sim 2$ and $\bp \sim 1$. 
At high redshift both $\bn$ and $\bp$ become significantly larger.

The nonlinear biasing relation can alternatively be parameterized
by a general power series:
\be
\delg = \sum_{n=0}^\infty { b_n \over {n !} } \delta^n .
\label{eq:taylor}
\ee
Since $g$ must average to zero, this power series can be written as
\be
\av{g|\del} = b_1\del + {1\over 2}b_2 (\del^2 -\sig^2)
            + {1\over 6}b_3 (\del^3 -S) +... ,
\label{eq:taylor3}
\ee
where $\sig^2\equiv\av{\del^2}$, $S\equiv\av{\del^3}$, etc.
This determines the constant term $b_0$.
The constraint at $-1$ provides another relation between the parameters.
Therefore, the expansion to third order contains only two free parameters
out of four.

In order to evaluate the parameters $\bh$ and $\bt$ for these nonlinear
toy models, we approximate the density PDF as a Gaussian, or alternatively
as log-normal in 
$\rho/\bar\rho=1+\del$ (\eg, Coles \& Jones 1991; Kofman \etal 1994):
\be
P(\del)= {1 \over [2\pi \ln(1+\sig^2)]^{1/2}}\  {1 \over (1+\del)}\
\exp \left(-{[\ln(1+\del)-\ln(1+\sig^2)^{-1/2}]^2 \over 2\ln(1+\sig^2)}\right).
\label{eq:lognormal}
\ee
The only free parameter is $\sig$. The skewness, for example, is
$S=2\sig^4 +\sig^6$, etc.

For nonlinear biasing that is described by Taylor expansion to third order
(\eq{taylor3}) and a Gaussian or log-normal density PDF, 
assuming $b_2\ll b_1$ and $\sig\ll 1$, one obtains 
\be
{\bt^2 \over \bh^2} \simeq 1 +{1\over2} \left({b_2 \over b_1}\right)^2 \sig^2 .
\ee
This is always larger than unity, but the deviation is small.
Alternatively, using the functional form of \eq{powerb}, with 
$\bn$ ranging from 1 to 5, $\bp$ ranging from 0.5 to 3, 
and a log-normal PDF of $\sig=0.7$, we find 
numerically that $\bt/\bh$ is in the range 1.0 to 1.15.
These two toy models, which approximate the nonlinear biasing behavior seen
for halos in the $N$-body simulations, indicate that despite the obvious
nonlinearity, especially in the negative regime, the nonlinear parameter
$\bt/\bh$ is typically only slightly larger than unity.
This means that the effects of nonlinear biasing on measurements of $\beta$
are likely to be relatively small.

\subsection{The Special Case of Gaussian Biasing}

A quick comment on ``Gaussian" and ``bivariate Gaussian" biasing, which
has been used in the recent literature (\eg, Pen 1998).

A specific model for nonlinear and stochastic biasing is where
the conditional distribution is a Gaussian, but allowing $b(\del)$ and
$\sb^2(\del)$ to vary with $\del$:
\be
P(g|\del) \propto \exp
   \left[-{[g-b(\del)\del]^2 \over 2\sig^2\sb^2(\del)} \right] .
\ee
Based on the $N$-body simulations, this is a reasonable approximation
for galactic halos.

However, the model of {\it bivariate Gaussian\,} biasing, 
which might be tempting
because it makes some of the computations easier, is much more
restrictive; it is in fact a special case of {\it linear\,} biasing.
This model assumes that the joint distribution of galaxies and mass is a
two-dimensional Gaussian,
\be
P(\tg,\td)\prop \exp \left[ -{\tg^2 -2r\tg\td +\td^2 \over 2(1-r^2)}
\right] ,
\ee
where $\tg\equiv g/\sigma_g$ and $\td\equiv \del/\sigma$,
and with the local biasing correlation coefficient $\cor=$const.

If $P(\del)$ is also a Gaussian, $P(\td)\prop \exp [-\td^2/2]$,
the conditional probability is
\be
P(\tg\,|\td)={P(\tg,\td) \over P(\td)}
          \propto \exp \left[ -{(\tg-r\td)^2 \over 2(1-r^2)} \right] .
\label{eq:condG}
\ee
This is a one-dimensional Gaussian for $\tg$, with mean $r\td$ and
variance $(1-r^2)$.
Back to the quantities $g$ and $\del$, the conditional mean is
$  \la g|\del\ra = r \bv\, \del $,
where $\bv=\sg/\sig$ as usual.
This is thus a special case of {\it linear\,} biasing, with a constant
linear biasing parameter $b_1 = r \bv$ independent of $\del$, as in
\eq{lin-sto}.
From the conditional variance of the Gaussian distribution
in \eq{condG}, the biasing scatter is also a constant independent of $\del$,
$  \sb^2 = \bv^2 (1-r^2) $.
Based on $N$-body simulations, linear biasing could be a poor approximation
even for galactic halos.
The whole biasing scheme is characterized in this case by only two
parameters, $\bv$ and $r$, or alternatively $b_1$ and $\sb$,
independent of $\del$.

\subsection{Observations}

Direct constraints on the local biasing field 
should, in principle, be provided
by the data themselves, of galaxy density (\eg, from redshift surveys)
versus mass density (\eg, from peculiar velocity surveys, or gravitational
lensing).
It is a bit early to deduce the nonlinear shape of $b(\del)$ from these data
because of the large errors that they involve at present.
However, we note a qualitative example of scatter in the biasing relation
in the fact that the smoothed density peaks
of the Great Attractor (GA) and Perseus Pisces (PP) are of comparable
height in the mass distribution as recovered by
POTENT from observed velocities (\eg, Dekel 1994; daCosta \etal 1996;
Dekel \etal 1998), but PP is significantly higher than GA in the galaxy map
(\eg, Hudson \etal 1995; Sigad \etal 1998).
For example,
a linear regression of the $12\hmpc$-smoothed density fields of POTENT
mass and optical galaxies in our cosmological neighborhood
yields a $\chi^2 \sim 2$ per degree of freedom
for the assumed errors in the data (Hudson \etal 1995).
One way to obtain a $\chi^2\sim 1$, as desired,
is to assume a biasing scatter of $\sb\sim 0.5$ in the optical density
(while $\sigma \sim 0.3$ at that smoothing).
With $b_1\sim 1$, one has $\sb^2/b_1^2 \sim 0.25$.
This is only a very crude estimate, and there is yet much to
be done along similar lines with future data.

\def\Cd{C}
\def\Cg{C_{\rm g}}
A promising method has been worked out (Sigad \& dekel 1998; see also
Dekel 1998b)
for recovering the mean biasing function $b(\delta)$ and its associated 
parameters $\bh$ and $\bt$ from a measured PDF (or counts in cells) of 
galaxies in a redshift survey.  If $g(\delta)$ were deterministic and 
monotonic, then it could be derived from the cumulative PDFs of
galaxies and mass, $\Cg(g)$ and $\Cd(\del)$, via
$g(\del) = \Cg^{-1} [ \Cd (\del)]$ (also Narayanan \& Weinberg 1998).
It is found for halos in $N$-body simulations that this
is a good approximation for $\av{g|\del}$ despite the scatter.
The other key point confirmed by a suite of simulations is that 
$\Cd(\del)$ is relatively insensitive to the cosmological model
or the fluctuation power spectrum, and can be approximated for our 
purpose by a log-normal distribution in $1+\del$ 
(\eg, Bernardeau 1994; Bernardeau \& Kofman 1995).
Thus, $b(\del)$ can be evaluated from a measured $\Cg(g)$ and the rms 
$\sigma$ of mass fluctuations on the same smoothing scale.
Since redshift surveys are by far richer than peculiar-velocity samples, 
this method will allow a better handle on $b(\del)$ than the local 
comparison of density fields of galaxies and mass. It can be applied 
to local redshift surveys as well as surveys of objects at high redshift.

Constraints on the biasing scheme can also be obtained by comparing the
clustering properties of galaxies of different types in a given
redshift surveys (\eg, Lahav \& Saslaw 1992).  
Indeed, partly motivated by the ideas of our current paper, 
a clear confirmation for nontrivial biasing,
non-linear and/or stochastic beyond shot noise, has recently been reported by 
Tegmark \& Bromley (1998) based on the Las Campanas Redshift Survey.

\section{CONCLUSIONS}
\label{sec:conc}

We have introduced a straightforward formalism for describing the
biasing relation between the density fluctuation fields of galaxies and
mass, based on the conditional probability function $P(\delg\vert\del)$.
The key feature of this formalism is the natural separation between
nonlinear and stochastic effects in the biasing scheme.
The nonlinearity is expressed by the conditional
mean via the function $b(\del)$, and the  
statistical scatter is measured by the
conditional standard deviation, $\sb(\del)$, and higher moments if necessary.
For analyses using local moments of second order, 
the biasing scheme is characterized by three parameters: 
$\bh$ measuring the mean biasing,
$\bt/\bh$ measuring the effect of nonlinearity, 
and $\sb/\bh$ measuring the effect of stochasticity.

Deviations from linear and deterministic biasing typically
result in biased estimates of the biasing parameter, or the parameter $\beta$
($\sim \Omega^{0.6}/b$),
which depend on the actual method of measurement.
The nonlinearity and the scatter lead to differences of order
$\bt^2/\bh^2$ and $\sb^2/\bh^2$ respectively in the different
estimators of $\beta$ using second-order local
moments.  They typically lead to an underestimate
of $\beta$ with respect to $\betah=f(\Omega)/\bh$.
Based on $N$-body simulations and toy models,
the effects of nonlinear biasing are typically small, on the order of 20\%
or less, and the effects of scatter could be somewhat larger.
One expects the $\beta$ parameters from second-order local moments 
to be biased in the following order: $\betainv < \betav < \betah$.
 
The stochasticity affects the redshift-distortion analysis only by
limiting the useful range of scales, especially in the analysis involving
power spectra. In this range, for {\it linear\,}
stochastic biasing, the basic {\it linear\,} expression reduces to the simple
Kaiser formula for $b(\del)=\bh=b_1$ (not $\bv$), and it does not involve the
scatter at all.
The distortion analysis is in principle sensitive to the
nonlinear properties of biasing, but the nonlinear effects, especially
at low redshifts, are expected to be weak, and on the same order as 
the effects of nonlinear gravitational instability. 
This is good news for the prospects of measuring an unbiased $\beta$ 
from redshift distortions in the large redshift surveys of the near 
future (2dF and SDSS).
A detailed nonlinear analysis of redshift distortions with nonlinear
biasing will be reported in a subsequent paper.

More detailed studies of simulations, including different recipes for galaxy
formation, are required in order to constrain the parameters of the biasing
formalism more accurately. The analysis could also be extended to include 
non-local biasing, using the biasing correlations as defined here. 
 
The study of stochastic and nonlinear biasing
should be extended to address the time evolution of biasing
because many relevant measurements of galaxy clustering
are now being done at high redshifts.
As seen in Fig.~\ref{fig:1},
the biasing is clearly a function of cosmological epoch
(\eg, Rees, private communication; Dekel \& Rees 1987;  
Mo \& White 1996; Steidel et al. 1996; 1998;
Bagla 1998a, 1998b; Matarrese \etal 1997; Wechsler \etal 1998; Peacock 1998).
In particular, if galaxy formation is limited to a given epoch
and the biasing is linear, one can show (\eg, Fry 1996) that
the linear biasing factor $b_1$ would eventually approach unity
as a simple result of the continuity equation.
Tegmark \& Peebles (1998) have generalized the analytic study of
time evolution to the case of stochastic but still linear biasing 
and showed how $\bv$ and $r$ approach unity in this case.
Analytic attempts to study the evolution of mildly nonlinear       
stochastic biasing have been reported recently 
(Taruya, Koyama \& Soda 1998; Taruya \& Soda 1998;
Catelan \etal 1998; Catelan, Matarrese \& Porciani 1998;
Sheth \& Lemson 1998).
These studies can be extended to the general nonlinear case using
our formalism.
The simulations of Somerville \etal (1998) are aimed at this goal.

More accurate measurements of peculiar velocities in our greater cosmological
neighborhood, and careful comparisons to the galaxy distribution, 
promise to allow improved observational estimates of the biasing scatter in
the future. The reconstruction of the large-scale mass distribution
based on weak gravitational lensing 
(Van Waerbeke 1998; 1999; Schneider 1998; Kaiser \etal 1998) 
is also becoming promising for this purpose.

The main moral of this paper is that in order to put any measurement 
of $\beta$ in cosmological perspective, and in particular when trying to
use it for an accurate measurement of the cosmological parameter $\Omega$,
one should consider the effects of nonlinear and stochastic biasing
and the associated complications of scale dependence, time dependence, and
type dependence.
The current different estimates are expected to span a range of 
$\sim 30\%$ in $\beta$ due to stochastic and nonlinear biasing.
The analysis of redshift distortions seems to be most promising;
once it is limited to the appropriate range of scales, the analysis is
independent of stochasticity and the nonlinear effects are expected to be 
relatively small.

\acknowledgments{ 
We thank Adi Nusser, Ue-Li Pen, Martin Rees, Michael Strauss, Max Tegmark,
and S.D.M. White for stimulating discussions.
This research was supported in part by the US-Israel Binational Science
Foundation grant 95-00330, and by the Israel Science Foundation grants
950/95 and 546/98.
}


\def\re{\reference}
\def\jeru{in {\it Formation of Structure in the Universe},
     eds.~A. Dekel \& J.P. Ostriker (Cambridge Univ. Press)\ }

\newpage


\begin{references}

\re{} Babul, A., \& White, S. D. M. 1991, \mnras, 253, L31
\re{} Bagla, J. S. 1998, \mnras, 297, 251                     
\re{} Bagla, J. S. 1998, \mnras, 299, 417                      
\re{bbks} Bardeen, J., Bond, J. R., Kaiser, N., \& Szalay, A.
    1986, \apj, 304, 15
\re{} Bartlett, J.B., \& Blanchard, A. 1996, \aap, 464, 805 
\re{ber94} Bernardeau, F. 1994, \aap, 291, 697                    
\re{ber95} Bernardeau, F., \& Kofman, L. 1995, \apj, 443, 479     
\re{} Blanton, M., Cen, R., Ostriker, J. P., \& Strauss, M. A.    
   1998, submitted (astro-ph/9807029)
\re{} Bond, J.R., Cole, S., Efstathiou, G., \& Kaiser, N. 1991, \apj,
   379, 440
\re{} Braun, E., Dekel, A., \& Shapiro, P. 1988, \apj, 328, 34
\re{beks} Broadhurst, T. J., Ellis, R. S., Koo, D. C.,
   \& Szalay, A. S., 1990, Nature, 343, 726
\re{} Catalan, P., Matarrese, S., \& Porciani, C. 1998, \apj, in press
     (astro-ph/9804250)                                          
\re{} Catalan, P., Lucchin, F., Matarrese, S., \& Porciani, C. 1998, 
    \mnras, 297, 692                       
\re{c1} Cen, R., \& Ostriker, J. P. 1992, ApJ, 399, L113
\re{c2} Cole, S., Fisher, K. B., \& Weinberg, D. 1995, \mnras, 275, 515
\re{} Coles, P., \& Jones, B. 1991, \mnras, 248, 1
\re{} da Costa, L. N., Freudling, W., Wegner, G., Giovanelli, R.,
   Haynes, M. P., \& Salzer, J. J. 1996, \apj , 468, L5
\re{} Davis, M., Efstathiou, G., Frenk, C. S., \& White, S. D. M. 1985,
    \apj, 292, 371
\re{} Davis, M., Miller, A., \& White, S. D. M. 1997, \apj, 460, 63 
\re{d7} Dekel, A. 1994, \araa, 32, 371
\re{} Dekel, A. 1997,
  in {\it Galaxy Scaling Relations: Origins, Evolution and 
  Applications\/} (ESO workshop, November 1996), ed.~L. da Costa 
  (Springer) in press (astro-ph/9705033)
\re{} Dekel, A. 1998a, \jeru in press
\re{} Dekel, A. 1998b, in Large-Scale Surveys (IAP Symposium XIV) 
   eds. Y. Mellier \& S. Colombi (Gif-sur-Yvette: Editions Fronti\`eres), 
   in press (astro-ph/9809291)
\re{d3} Dekel, A., Bertschinger, E., Yahil, A., Strauss, M. A., 
  Davis, M., \& Huchra J. P. 1993, \apj , 412, 1 (PI93)
\re{dd} Dekel, A., Burstein, D., \& White, S.D.M.
   1997, in {\it Critical Dialogues in Cosmology}, ed. N. Turok
   (Singapore: World Scientific), p. 175 (astro-ph/9611108)
\re{d1} Dekel, A., Eldar, A., Kolatt, T., Willick, J. A., Faber,
   S. M., Courteau, S., \& Burstein, D. 1998, in preparation
\re{} Dekel, A., \& Rees, M. J. 1987, \nat, 326, 455
\re{} Dekel, A., \& Silk, J. 1986, \apj, 303, 39
\re{d5} Dressler, A. 1980, \apj, 236, 351
\re{} Fisher, K. B., Davis, M., Strauss, M. A., Yahil, A., \&
   Huchra, J. P. 1994, \mnras, 267, 927 
\re{} Fisher, K. B., \& Nusser, A. 1996, \mnras, 279, L1 
\re{} Fisher, K. B., Scharf, C. A., \& Lahav, O. 1994, \mnras, 266, 219
\re{} Fry, J. N. 1996, \apjl, 461, L65
\re{} Guzzo, L., Strauss, M. A., Fisher, K. B., Giovanelli, R., 
    \& Haynes, M. P.   1997, \apj, 489, 37
\re{hamil92} Hamilton, A. J. S. 1992, \apjl, 385, L5  
\re{hamil93} Hamilton, A. J. S. 1993, \apjl, 406, L47  
\re{} Hamilton, A. J. S. 1995, in {\it Clustering in the Universe},
   Proc. 30$^{\rm th}$ Rencontres de Moriond, ed.~S. Maurogordato,
   C. Balkowski, C. Tao, and J. Tr\^an Thanh V\^an (Gif-sur-Yvette:
   Editions Fronti\`eres), 143
\re{} Hamilton, A. J. S. 1997, \mnras, 289, 285
\re{} Heavens, A. F., \& Taylor, A. N. 1995, \mnras, 275, 483
\re{} Hermit, S., Santiago, B. X., Lahav, O., Strauss, M. A., Davis, M., 
     Dressler, A., \& Huchra, J. P. 1996, \mnras, 283, 709
\re{h3} Hudson, M. J., Dekel, A., Courteau, S., Faber,
   S. M., \& Willick, J. A. 1995, \mnras , 274, 305
\re{} Jing, Y. P., \& Suto, Y. 1998, \apjl, 494, L5
\re{} Juszkiewicz, R., Bouchet, F. R., \& Colombi, S. 1993, \apjl, 412 L9
\re{} Kaiser, N. 1984, \apjl, 284, L9
\re{} Kaiser, N. 1987, \mnras , 227, 1
\re{} Kaiser, N., Wilson, G., Luppino, G., Kofman, L., Gioia, I.,
      Metzger, M., \& Dahle, H. 1998, (astro-ph/9809268)
\re{k4} Kauffman, G., Nusser, A., \& Steinmetz, M. 1997, \mnras, 286, 795
\re{} Kirshner, R. P., Oemler, A. Jr., Schechter, P. L., 
   \& Shectman, S. A. 1987 \apj, 314, 493
\ref{} Kofman, L., Bertschinger, E., Gelb, J., Nusser, A. and Dekel, A.
  1994, \apj , 420, 44
\re{} Kolatt, T., \& Dekel, A. 1997, \apj, 479, 592
\re{} Lahav, O. 1996, Helvetica Physica Acta, 96, 388  
\re{} Lahav, O., Nemiroff, R. J., \& Piran, T. 1990, \apj, 350, 119
\re{} Lahav, O., \& Saslaw, W. 1992, \apj, 396, 430
\re{} Loveday \etal 1995, \apj , 442, 457
\re{} Matarrese, S., Coles, P., Lucchin, F., \& Moscardini, L. 1997, 
      \mnras, 286, 115
\re{m3} Mo, H., \& White, S.D.M. 1996, \mnras , 282, 347
\re{nar} Narayanan, V. K., \& Weinberg, D. H. 1998, \apj, in press
      (astro-ph/9806238)
\re{n3} Nusser, A., Dekel, A., Bertschinger, E., \& Blumenthal,
   G. R. 1991, \apj, 379, 6
\re{} Peacock, J. A. 1998, Phil. Trans. R. Soc. Lond. A, submitted 
   (astro-ph/9805208).
\re{} Peebles, P. J. E. 1980, {Large-Scale Structure in the Universe}
    (Princeton University Press)
\re{} Peebles, P. J. E. 1993, {Principles of Physical Cosmology} (Princeton
    University Press) 
\re{} Pen, U.-L. 1998, ApJ, 504, 601
\re{} Santiago, B. X. \& Strauss, M. A. 1992, \apj, 387, 9
\re{} Scherrer, R.J. \& Weinberg, D.H. 1998, ApJ, 504, 607
\re{} Schneider, P. 1998, ApJ, 498, 43
\re{} Sheth, R. K., \& Lemson, G. 1998, MNRAS, submitted 
      (astro-ph/9808138)                    
\re{} Sigad, Y., \& Dekel, A. 1998, in preparation
\re{} Sigad, Y., Eldar, A., Dekel, A., Strauss, M. A., \& Yahil, A.
   1998, \apj, 495, 516 (astro-ph/9708141)
\re{} Somerville, R., Sigad, Y., Lemson, G., Dekel, A., 
      Colberg, J., Kauffmann, G., \& White, S. D. M. 1998, in preparation
\re{} Steidel, C. C., Adelberger, K. L., Dickinson, M., Giavalisco, M., 
   Pettini, M., \& Kellogg, M. 1998, \apj, 492, 428 
\re{} Steidel, C. C., Giavalisco, M., Pettini, M., Dickinson, M., \&
   Adelberger, K. L. 1996, \apjl, 462, L17
\re{s2}Strauss, M. A., \& Willick, J. A. 1995, \physrep, 261, 271
\re{} Szapudi, I. 1998, \mnras, submitted (astro-ph/9805090)     
\re{} Taruya, A., Koyama, K., \& Soda, J. 1998, submitted
      (astro-ph/9807005)                                         
\re{} Taruya, A., \& Soda, J. 1998, submitted (astro-ph/9809204) 
\re{} Tegmark, M. \& Bromley, B. C. 1998, submitted (astro-ph/9809324) 
\re{} Tegmark, M. \& Peebles, P. J. E. 1998, submitted (astro-ph/9804067)
\re{} Van Waerbeke, L. 1998, A\&A, 334, 1                        
\re{} Van Waerbeke, L. 1999, (astro-ph/9807041)                  
\re{} Wechsler, R. H., Gross, M. A. K., Primack, J. R., Blumenthal, G. R.,
   \& Dekel, A. 1998, \apj, 506, 19 (astro-ph/9712141)
\re{} Weinberg, D. H. 1995, in {Wide-Field Spectroscopy and the
    Distant Universe}, eds. S. J. Maddox \& A. Aragon-Salamanaca 
    (Singapore: World Scientific), 129

\end{references}
\end{document}